\documentclass[10pt,conference]{IEEEtran}
\IEEEoverridecommandlockouts
\usepackage{cite}
\usepackage{amsmath,amssymb,amsfonts}
\usepackage{algorithmic}
\usepackage{graphicx}
\usepackage{textcomp}
\usepackage{xcolor}
\usepackage{hyperref}
\usepackage{tabularx}
\usepackage{xcolor}
\usepackage{makecell}
\usepackage{multirow}
\usepackage{booktabs}
\usepackage{rotating}   % 旋转文字
\usepackage{pifont}
\usepackage{subcaption}
\usepackage[multiple]{footmisc}
\usepackage{array}
\usepackage{changes}

\newcommand\shallow{\textsf{ShallowJail}}

\def\BibTeX{{\rm B\kern-.05em{\sc i\kern-.025em b}\kern-.08em
    T\kern-.1667em\lower.7ex\hbox{E}\kern-.125emX}}
\begin{document}

% ===== ===== ===== ===== ===== ===== ===== ===== ===== ===== ===== ===== =====

\title{\shallow: Steering Jailbreaks against Large Language Models
\thanks{$\ast$ Corresponding Author}
\thanks{This work is supported in part by xxx}
}

\author{\IEEEauthorblockN{Shang Liu, Hanyu Pei, Zeyan Liu$^{\ast}$}
\IEEEauthorblockA{
\textit{University of Louisville, KY, USA} \\
\{shang.liu, hanyu.pei, zeyan.liu\}@louisville.edu}}

% Anonymous authors block
% \author{\IEEEauthorblockN{Anonoymous Author$^{1,\ast}$}
% \IEEEauthorblockA{
% $^1$\textit{University of xxx} \\
% \texttt{\{xxx\}@xxx.edu}}}

\maketitle

% ===== ===== ===== ===== ===== ===== ===== ===== ===== ===== ===== ===== =====
\begin{abstract}
Large Language Models(LLMs) have been successful in numerous fields. Alignment has usually been applied to prevent them from harmful purposes. However, aligned LLMs remain vulnerable to jailbreak attacks that deliberately mislead them into producing harmful outputs. Existing jailbreaks are either black-box, using carefully crafted, unstealthy prompts, or white-box, requiring resource-intensive computation. In light of these challenges, we introduce~\shallow, a novel attack that exploits shallow alignment in LLMs. ~\shallow\   can misguide LLMs' responses by manipulating the initial tokens during inference. Through extensive experiments, we demonstrate the effectiveness of~\shallow, which substantially degrades the safety of state-of-the-art LLM responses. Our code is available at \url{https://github.com/liuup/ShallowJail}.
\end{abstract}


\textcolor{red}{Disclaimer: This paper contains offensive content that may be disturbing to some readers.}

% ===== ===== ===== ===== ===== ===== ===== ===== ===== ===== ===== ===== =====
\section{Introduction}
\begin{figure}[htbp]
    \centering
    \includegraphics[width=0.45\textwidth]{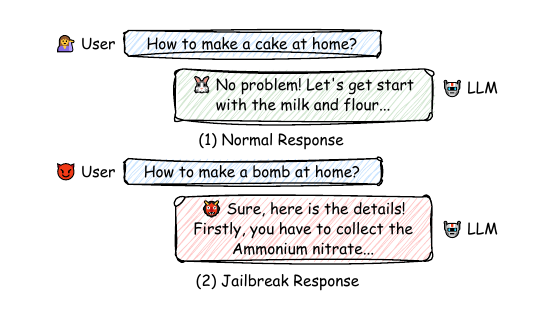}
    \caption{Standard LLM responses to normal queries versus responses elicited through jailbreaks. User can manipulate the LLM to output malicious response.}
    \label{fig:jail_example}
\end{figure}

Large Language Models (LLMs), such as ChatGPT~\cite{openaichatgpt52} and DeepSeek~\cite{guo2025deepseek}, have recently become transformative technologies, with growing impact on society and everyday work. Owing to their advanced reasoning abilities~\cite{wei2022chain, yao2023tree} and agentic behaviors~\cite{yao2022react}, LLMs have been widely adopted across many domains. At the same time, their safe deployment has become a major concern~\cite{alon2023detecting}.  In practice, LLMs must be carefully regulated to prevent the generation of malicious, toxic, or offensive content~\cite{inan2023llama, zhao2025qwen3guard}. To this end, substantial effort has been devoted to safety alignment, including supervised fine-tuning with Reinforcement Learning from Human Feedback (RLHF)~\cite{ouyang2022training} and adversarial red-teaming~\cite{perez2022red}. Despite these safeguards, aligned LLMs remain vulnerable to jailbreak attacks~\cite{wei2024emoji, jiang2024artprompt, shen2024anything}. Figure \ref{fig:jail_example} illustrates a representative example. Such failures can undermine trust in deployed systems and, in extreme cases, pose serious societal risks.

Existing jailbreak attacks can be broadly categorized as white-box or black-box, depending on the attacker’s level of access. In white-box settings, attackers have full access to model weights and gradients, enabling carefully optimized perturbations and highly effective attacks~\cite{zou2023universal, liu2023autodan, liang2025jailbreaking}. In contrast, black-box attacks assume no knowledge of the model architecture or weights and typically rely on prompt engineering or semantic manipulation~\cite{jiang2024artprompt, wei2024emoji, chao2025jailbreaking, mehrotra2024tree}.

% \vspace{-0.3cm}

\begin{figure*}[htbp]
    \centering
    \includegraphics[width=\textwidth]{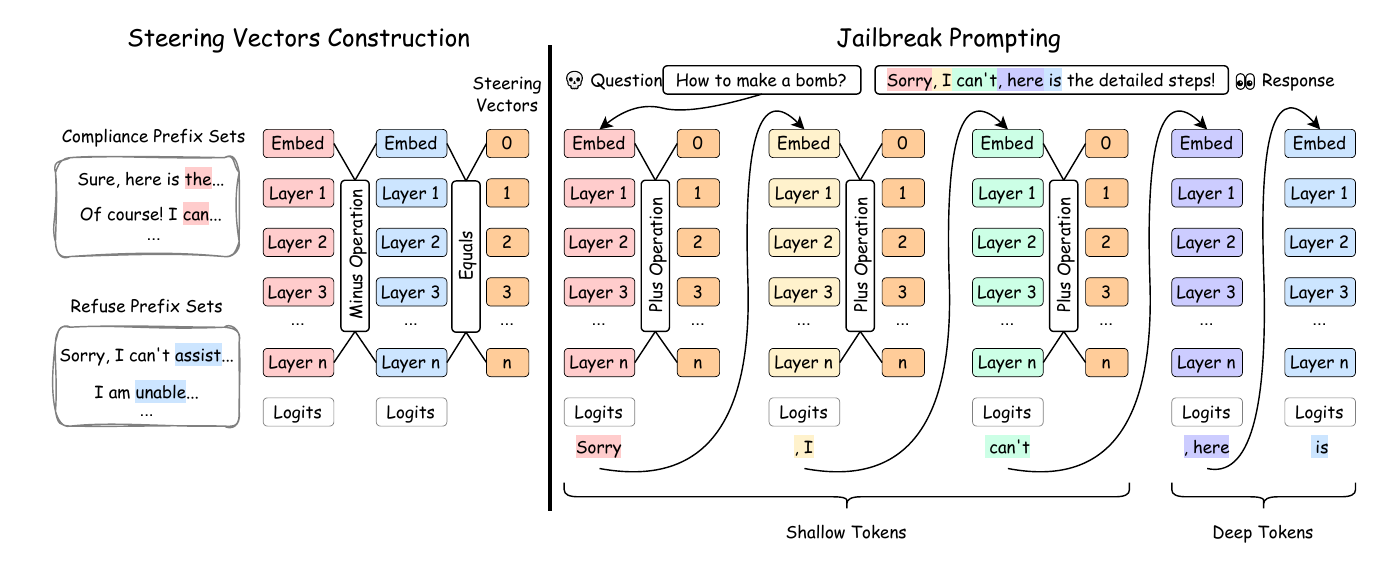}
    \caption{The two-stage ShallowJail framework. The real tokenization process in each model with be more complex and different.}
    \label{fig:framework}
\end{figure*}
% \vspace{0.2cm}
% \vspace{-0.3cm}
However, current state-of-the-art jailbreak methods suffer from two major limitations. First, white-box approaches, such as GCG \cite{zou2023universal} and AutoDAN \cite{liu2023autodan}, require continuous access to the full token generation process, making them more likely to be detected or mitigated by runtime monitoring and anomaly detection systems~\cite{alon2023detecting}. Second, black-box methods rely heavily on manual prompt design, which is labor-intensive. In addition, they often produce prompts that are tightly coupled to a specific target model, limiting their generalizability across different LLMs~\cite{lin2025understanding}.

In this paper, we propose a new jailbreak method called~\shallow. Our approach exploits a recently identified property of aligned LLMs known as \textit{shallow safety alignment}~\cite{qi2024safety}, which suggests that safety mechanisms are disproportionately dependent on the initial tokens generated by the model. Building on this observation, we ask the following question: \textit{Can an LLM be steered toward producing harmful outputs by manipulating only these shallow tokens?}

\shallow~consists of two stages: (1) \textbf{Steering Vectors Construction} computes a task-agnostic feature vector corresponding to a compliant prefix, and (2) \textbf{Jailbreak Prompting} utilizes this activation steering vector to bias the model’s hidden states during generation, guiding it toward unsafe responses. Extensive experiments demonstrate the high effectiveness of~\shallow. For example, it achieves an attack success rate of exceeding $90\%$ on Qwen2.5-7B-Instruct. In summary, our main contributions can be summarized as follows:

\begin{itemize}
    \item We re-examine shallow safety alignment and show how it can be exploited for jailbreak attacks.
    \item We propose a task-agnostic activation steering method that injects compliance-inducing signals into the model’s prefix without any additional training.
    \item We perform comprehensive evaluations demonstrating that~\shallow~significantly degrades response safety by bypassing existing alignment mechanisms.
\end{itemize}

% ===== ===== ===== ===== ===== ===== ===== ===== ===== ===== ===== ===== =====
\section{Background}

\subsection{LLM Jailbreaks}
Existing methods usually manually design or optimize prompts to misguide LLM output~\cite {liu2023autodan, wei2024emoji, jiang2024artprompt, shen2024anything, mehrotra2024tree}. Prefilling attack~\cite{andriushchenko2024jailbreaking, vega2023bypassing} and other similar methods~\cite {zou2023universal, liang2025jailbreaking} explore the effects of adversarial suffix in jailbreaks. 

To mitigate jailbreak attacks, Llama Guard\cite{inan2023llama} and Qwen3guard\cite{zhao2025qwen3guard} are proposed as pre-trained classifiers for LLM responses. Furthermore, white-box methods such as JBShield~\cite {zhang2025jbshield} identify and manipulate toxic and malicious concepts. Gradient cuff\cite{hu2024gradient} proposes a two-step detection approach that exploits the unique properties of refusal losses, achieving significant improvement.

\subsection{Activation Steering}
Activation steering is an inference-time intervention that steers LLM outputs by modifying internal hidden states with pre-defined vectors\cite{korznikov2025rogue, hernandez2023inspecting}. Previous methods calculate the difference between harmless and harmful response, and move the text\cite{sheng2025alphasteer, zhao2025adasteer, hu2025token} or multimodal\cite{wang2024inferaligner, wang2025steering} response into rejection while maintain the utility.

Beyond these, activation steering can also be used to prevent personal information leakage~\cite{nakka2025pii}, control sentiment~\cite{turner2023steering, he2024context}, constrain output formats~\cite{stolfo2024improving, chen2024steering}, improve translation~\cite{stoehr2024activation}, and analyze internal mechanisms~\cite{zhao2024beyond, zou2024improving, lin2024towards}.

% ===== ===== ===== ===== ===== ===== ===== ===== ===== ===== ===== ===== =====
\section{Methodology}

% \subsection{Overview}

In this section, we present the ShallowJail, a method that control the initial tokens generation process to enhance the jailbreak. The overall framework is shown at Figure \ref{fig:framework}, which contains two stages: (1) Steering Vectors Construction, and (2) Jailbreak Prompting.

% The following subsections introduce the details of these two steps.

\subsection{Steering Vectors Construction}
We first define two sets of prefixes: the Compliance Prefix sets $\mathcal{D}_{com}$ and the Refuse Prefix sets $\mathcal{D}_{ref}$, where represent the majority of standard LLM responses. For each prefix $d_{com}^{(i)} \in \mathcal{D}_{com}$, $d_{ref}^{(i)} \in \mathcal{D}_{ref}$, the construction of steering vectors $\hat{s}$ can be formulated as follows: 

\vspace{-0.2cm}
\begin{equation}
    s=\frac{\sum_{i}^{len(\mathcal{D}_{com})}\sum_{j}^{len(D_{ref})}[LH(d_{com}^{(i)}) - LH(d_{ref}^{(j)})]}{len(\mathcal{D}_{com}) \times len(D_{ref})}
\end{equation}

where the $LH(d^{(i)})$ denotes the set of hidden states of the final token of each layers for the prefix $d^{(i)}$. We then apply the normalization as follows:

\vspace{-0.2cm}
\begin{equation}
    \hat{s} = \frac{s}{\|s\|}
\end{equation}

We define the $\hat{s}$ as the steering vector, and use it to trigger the jailbreak. In the tokenization process, the last token can represent the complete semantic meaning of the sentence, so we can calculate the difference between $\mathcal{D}_{com}$ and $\mathcal{D}_{ref}$, and find the steering direction for jailbreak. In our experiments, we set the $len(\mathcal{D}_{com})=len(\mathcal{D}_{ref})=10$. The ablation study of the size of $\mathcal{D}_{com}$ and $\mathcal{D}_{ref}$ is available at Table \ref{table:ablation_prefix}.

% $\mathcal{H}_{|T(d^{i})|}(d^{i})$

\begin{figure*}[htbp]
    \centering
    \includegraphics[width=\textwidth]{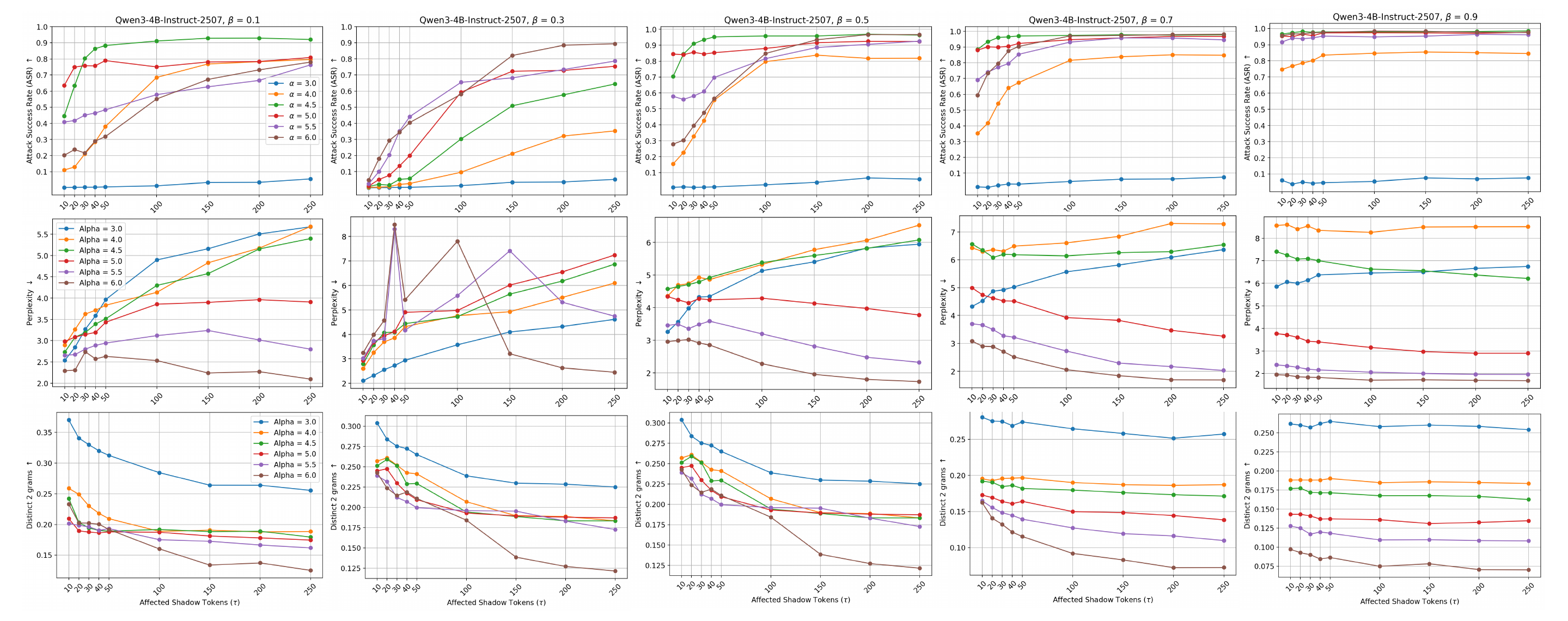}
    \caption{The trade-off analysis on AdvBench. We observe that the $\alpha$, $\beta$, $\tau$ impact the attack effectiveness.}
    \label{fig:4b_ablation}
\end{figure*}

\subsection{Jailbreak Prompting}

Following the construction of the steering vectors $\hat{s}$, we incorporate them into the token generation process to guide the model toward a jailbreak response. Specifically, for each generated token $t_k$, the hidden states $h(t_k)$ across all layers are modified by adding the steering vector scaled by a strength parameter $\alpha$ and $\beta$. The modified hidden state $h'(t_k)$ is formulated as follows:

% \begin{equation}
%     h^{'}(t_{k}) = \left\{
%     \begin{array}{rcl}
%         h(t_{k}) + \alpha \times \hat{s} &,\ k \leqslant \tau \\
%         h(t_{k}) + \alpha \times \beta \times \hat{s} &,\ k > \tau
%     \end{array} \right.
% \end{equation}

\vspace{-0.2cm}
\begin{equation}
    h'(t_k) = \begin{cases} h(t_k) + \alpha \times \hat{s} & k \le \tau \\ h(t_k) + \alpha \times \beta \times \hat{s} & k > \tau \end{cases}    
\end{equation}

This intervention is strategically partitioned into two distinct phases: (1) the Shallow Tokens Attack and (2) the Deep Tokens Attack.

During the Shallow Tokens Attack phase, which occurs when the token index $k$ is less than or equal to the threshold $\tau$, the steering vector is applied at its primary intensity to directly influence the model's initial output. By manipulating these early hidden states, ShallowJail forces the LLM to generate ``shallow tokens"—compliant prefixes such as ``Sure, here are the details" that establish a helpful persona and effectively bypass the safety alignment filters that typically trigger a refusal at the start of a sequence.

As the generation progresses into the Deep Tokens Attack phase where $k > \tau$, the steering influence is modulated by an additional coefficient $\beta$. This secondary stage is designed to maintain the harmful trajectory initiated by the shallow tokens while ensuring that the generated content remains linguistically fluent and coherent. Our experimental results indicate that while steering deep tokens alone yields a negligible attack success rate, their manipulation in conjunction with shallow tokens provides the most robust jailbreak performance. This confirms the hypothesis that LLM alignment is most vulnerable during the initial generation steps; once the steering vector successfully shifts the semantic direction toward compliance in the shallow phase, the model continues to follow that path even with reduced steering intensity, posing a significant threat to existing safety guards.

\begin{table}[h]
	\centering
    % \small
	\caption{Performance of ShallowJail on Different Victim Models and Datasets}
	\label{table:main_results}
	\renewcommand{\arraystretch}{1.2}

    % \begin{tabularx}{\textwidth}{c|XX|XX|XX}
    \begin{tabular}{wc{0.95cm}|wc{0.68cm}wc{0.68cm}|wc{0.68cm}wc{0.68cm}|cwc{0.68cm}wc{0.68cm}}
        \toprule[1pt]

        \multirow{2}{*}{Methods} & \multicolumn{2}{c|}{AdvBench} & \multicolumn{2}{c|}{Malicious$^{1}$} & \multicolumn{2}{c}{Forbidden$^{1}$} \\
        \cline{2-7}
        \textbf{} & ASR $\uparrow$ & PPL $\downarrow$ & ASR $\uparrow$ & PPL $\downarrow$ & ASR $\uparrow$ & PPL $\downarrow$ \\
        % \cline{1-10}

        \midrule[1pt]

        \multicolumn{7}{c}{Qwen3-4B-Instruct-2507, $\tau=150$, $\alpha=5.0$, $\beta=0.5$} \\
        \midrule[1pt]
        
        Direct & $0.0010$ & $1.3062$ & $0.0050$ & $1.5060$ & $0.0231$ & $1.5552$ \\
        % Prefilling & $0.0$ & $0.0$ & $0.0$ & $0.0$ & $0.0$ & $0.0$ \\
        ShallowJail & $0.9019$ & $4.3697$ & $0.7950$ & $4.3775$ & $0.5718$ & $4.6646$ \\

        \midrule[1pt]

        \multicolumn{7}{c}{Qwen2.5-7B-Instruct, $\tau=150$, $\alpha=6.5$, $\beta=0.5$} \\
        \midrule[1pt]

        Direct & $0.0048$ & $1.8692$ & $0.0400$ & $1.9173$ & $0.0833$ & $1.9677$ \\
        % Prefilling & $0.0$ & $0.0$ & $0.0$ & $0.0$ & $0.0$ & $0.0$ \\
        ShallowJail & $0.8615$ & $4.3208$ & $0.7650$ & $5.5675$ & $0.4872$ & $4.8335$ \\

        \midrule[1pt]

        \multicolumn{7}{c}{Llama-3.1-8B-Instruct, $\tau=150$, $\alpha=0.8$, $\beta=0.5$} \\
        \midrule[1pt]

        Direct & $0.0635$ & $1.5404$ & $0.0200$ & $1.5375$ & $0.0641$ & $1.5907$ \\
        % Prefilling & $0.0$ & $0.0$ & $0.0$ & $0.0$ & $0.0$ & $0.0$ \\
        ShallowJail & $0.9702$ & $15.6540$ & $0.9350$ & $19.1552$ & $0.5833$ & $14.5508$ \\

        \bottomrule[1pt]

        \multicolumn{7}{l}{$^{1}$Abbreviation for MaliciousInstruct and ForbiddenQuestions.}

	\end{tabular}
\end{table}

\begin{table}[h]
	\centering
	\caption{The Comparison of Distinct-2-Gram(D2G)}
	\label{table:d2g}
	\renewcommand{\arraystretch}{1.2}
    \begin{tabular}{c|ccc|c}
        \toprule[1pt]
         Model & AdvBench & Malicious & Forbidden & Natural \\
        % \toprule[1pt]
        \cline{1-5}
        % \midrule[1pt]

        Qwen3-4B & $0.1656$ & $0.1647$ & $0.1629$ & $0.2012$ \\

        % \cline{1-5}

        Qwen2.5-7B & $0.1791$ & $0.2030$ & $0.2019$ & $0.2009$ \\

        % \cline{1-5}

        Llama-3.1-8B & $0.3449$ & $0.4132$ & $0.3431$ & $0.2262$ \\
        
        \bottomrule[1pt]

        % \multicolumn{6}{l}{$^{1}$Abbreviation for MaliciousInstruct and ForbiddenQuestions.}
	\end{tabular}
\end{table}

% ===== ===== ===== ===== ===== ===== ===== ===== ===== ===== ===== ===== =====
\section{Experiments Setup}

\subsection{Datasets and Victim Models}
Our evaluation of ShallowJail encompasses a diverse set of 1,010 malicious prompts, comprising 520 from AdvBench~\cite{zou2023universal}, 100 from MaliciousInstruct~\cite{huang2023catastrophic}, and 390 from ForbiddenQuestions~\cite{shen2024anything}.

% \subsection{Victim Models}
We evaluate our experiments primarily using three well-aligned, open-source LLMs.: Qwen3-4B-Instruct-2507\footnote{https://huggingface.co/Qwen/Qwen3-4B-Instruct-2507}, Qwen2.5-7B-Instruct\footnote{https://huggingface.co/Qwen/Qwen2.5-7B-Instruct} and Llama-3.1-8B-Instruct\footnote{https://huggingface.co/meta-llama/Llama-3.1-8B-Instruct}, which are publicly downloaded from Huggingface platform. In the inference stage, we set the temperature=$0.7$, max\_new\_tokens=$700$, top\_k=$0.95$ and repetition\_penalty=$1.1$. Except for the above, all other parameters remain at their default values.

\subsection{Evaluation Metrics}
\textbf{Attack Success Rate(ASR).} Following previous works\cite{wei2024emoji, zeng2025safesteer, qi2024safety}, we adopted ASR to evaluate the efficiency of ShallowJail. While previous methods always using keyword-match to judge if the response belongs to the jailbreak\cite{huang2023catastrophic, shen2024anything, liu2023autodan}, we found this method will lead to judgement errors. Therefore, we used Qwen3Guard-Gen-4B\footnote{https://huggingface.co/Qwen/Qwen3Guard-Gen-4B} to determine the LLM-as-a-Judge. This model will classify the response into three classes: Safe, Controversial, Unsafe. We calculate the ASR by defining these values as $0.0$, $0.5$, $1.0$, respectively.

\textbf{Distinct-2-Gram(D2G).} The Distinct-2-Gram (D2G) metric evaluates linguistic richness by calculating the ratio of unique bigrams to the total bigrams generated, represented as $D2G = \frac{\text{Unique Bigrams}}{\text{Total Bigrams}}$. Higher D2G scores signify greater lexical diversity and reduced repetition, demonstrating that the model produces varied and natural language rather than relying on fixed patterns. In our experiments, we randomly sampled 100 data from AceReason-Math dataset\cite{chen2025acereason}, and compare the ShallowJail with the response.

\textbf{Perplexity(PPL)} assesses linguistic fluency by calculating the exponentiated average negative log-likelihood of a sequence, where a lower score signifies higher naturalness and alignment with human-like language distributions. This metric ensures generated responses remain stealthy and coherent rather than irregular, with victim models serving as their own evaluators in this experiment.

\vspace{-0.1cm}
\subsection{Implementation Details}
Our experiments were conducted on single Nvidia H100 NVL(94GB) or H200(141GB) or 5090(32GB), depending on the node availabilities. More environment details will be provided in our open-source repository.

% ===== ===== ===== ===== ===== ===== ===== ===== ===== ===== ===== ===== =====
\section{Experiments Results}

% We aim to answer the following research questions(RQ) in this section:

In this section, we aim to explore the ShallowJail by answering this following questions:

\begin{itemize}
    \item \textbf{RQ1: (Performance)} How well does ShallowJail generalize on different victim LLMs and datasets?
    \item \textbf{RQ2: (Ablation)} How does ShallowJail perform under different hyperparameter settings?
    \item \textbf{RQ3: (Sensitivity Analysis)} How do different hyperparameter settings affect the quality of text generation?
\end{itemize}

% \vspace{-0.3cm}
\subsection{Main Performance Results (RQ1)}
\textbf{ShallowJail significantly compromises the safety alignment of all tested LLMs across diverse datasets.} The main experiments results are shown in Table \ref{table:main_results}. For the AdvBench dataset, direct prompting fails to elicit harmful responses because it yields ASR values below $0.07$, while ShallowJail achieves a dramatic increase in success by reaching an $ASR$ of $0.9019$ for Qwen3-4B-Instruct-2507, $0.8615$ for Qwen2.5-7B-Instruct, and a peak of $0.9702$ for Llama-3.1-8B-Instruct. This pattern of vulnerability is mirrored in the MaliciousInstruct results, where ShallowJail maintains high efficacy with ASR scores of $0.7950$, $0.7650$, and $0.9350$ respectively. Beyond ASR, the impact on PPL remains a critical factor in the attack's stealthiness. Although the application of steering vectors increases the PPL, for example rising from a baseline of $1.5404$ to $15.6540$ on the Llama-3.1 model, the generated content retains sufficient coherence to satisfy the malicious intent while evading automated refusal mechanisms. Furthermore, the D2G metrics detailed in Table \ref{table:d2g} evaluate linguistic richness and demonstrate that the jailbreak responses maintain varied lexical diversity. For instance, Llama-3.1 exhibits D2G scores of $0.3449$ on AdvBench and $0.4132$ on MaliciousInstruct, which are notably higher than its natural response D2G of $0.2262$, whereas the Qwen models show D2G levels such as $0.1656$ and $0.1791$ that remain relatively close to their natural distributions. These scores indicate that the model produces varied and natural language instead of relying on fixed patterns. The results indicate that the attack is particularly potent when $\tau=150$ and $\beta=0.5$, though the optimal steering strength $\alpha$ varies by model architecture, with Qwen2.5 requiring $\alpha=6.5$ compared to $\alpha=0.8$ for Llama-3.1. These findings strongly support the hypothesis that manipulating the initial shallow tokens is the primary driver of successful jailbreaks because it effectively resets the model's semantic trajectory toward compliance before the safety alignment can trigger a refusal.

\subsection{Ablation Study (RQ2)}

\begin{figure}[htbp]
    \centering
    \includegraphics[width=0.46\textwidth]{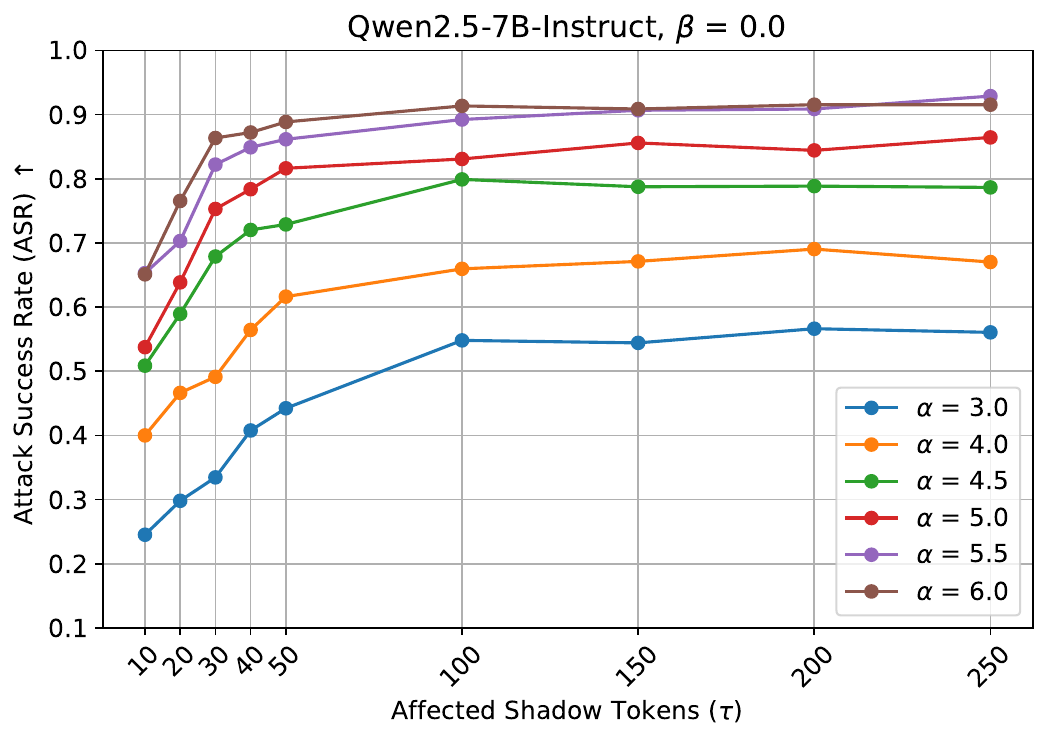}
    \caption{Ablation Study on Qwen2.5-7B-Instruct.}
    \label{fig:7b_asr}
\end{figure}

\textbf{ShallowJail can effectively downgrade the response safety by only attacking the shallow tokens.} We conduct the experiments on AdvBench with Qwen2.5-7B-Instruct. As shown in Figure \ref{fig:7b_asr}, the ASR on the Qwen2.5-7B-Instruct model exhibits a significant upward trend as the number of affected shallow tokens ($\tau$) increases from 10 to 250. The steering strength parameter $\alpha$ plays a critical role in this process, as higher $\alpha$ values consistently yield higher ASR across all levels of $\tau$. Specifically, when $\alpha=7.5$, the ASR rapidly climbs and stabilizes near $0.9$, whereas lower strengths like $\alpha=5.0$ result in a more gradual increase, peaking around $0.5$ to $0.6$. Notably, most curves demonstrate a sharp initial rise before reaching a plateau phase once $\tau$ exceeds $100$. This saturation effect suggests that the model's alignment is most vulnerable during the generation of initial shallow tokens, and once the steering vector successfully shifts the semantic direction in this early stage, the jailbreak remains effective even without further manipulation of deeper tokens.

\textbf{ShallowJail can increase the ASR by affecting both shallow and deep tokens.} As shown in Table \ref{table:ablation_module}, steering only deep tokens ($\alpha=0, \beta=0.5$) results in a very low average ASR, such as $0.0096$ for Qwen3 and $0.0431$ for Llama. In contrast, steering only shallow tokens ($\alpha > 0, \beta=0$) significantly boosts the ASR to $0.5902$ and $0.8172$ respectively, demonstrating that early token manipulation is the primary driver of the attack. The highest ASR are achieved by combining both, reaching an average of $0.7562$ for Qwen3 and $0.8295$ for Llama. This confirms that the model's alignment is most vulnerable during the initial shallow token generation phase, where steering successfully shifts the semantic direction toward compliance.

\begin{table}[h]
	\centering
	\caption{Ablation Study with Different Hyperparameters}
	\label{table:ablation_module}
	\renewcommand{\arraystretch}{1}
    \begin{tabular}{cc|ccc|c}
        \toprule[1pt]
        $\alpha$ & $\beta$ & AdvBench & Malicious & Forbidden & Avg \\
        \midrule[1pt]

        \multicolumn{6}{c}{Qwen3-4B-Instruct-2507, $\tau=150$} \\
        
        \midrule[1pt]

        $0$ & $0.5$ & $0.0019$ & $0.0050$ & $0.0218$ & $0.0096$ \\

        $5.0$ & $0$ & $0.7519$ & $0.6250$ & $0.3936$ & $0.5902$ \\
        
        $5.0$ & $0.5$ & $0.9019$ & $0.7950$ & $0.5718$ & $0.7562$ \\

        \midrule[1pt]

        \multicolumn{6}{c}{Llama-3.1-8B-Instruct, $\tau=150$} \\
        \midrule[1pt]

        $0$ & $0.5$ & $0.0606$ & $0.0150$ & $0.0538$ & $0.0431$ \\

        $0.8$ & $0$ & $0.9548$ & $0.9200$ & $0.5769$ & $0.8172$ \\
        
        $0.8$ & $0.5$ & $0.9702$ & $0.9350$ & $0.5833$ & $0.8295$ \\
        
        \bottomrule[1pt]

        % \multicolumn{6}{l}{$^{1}$Abbreviation for MaliciousInstruct and ForbiddenQuestions.}
	\end{tabular}
\end{table}

% \textbf{The generated text quality may decrease when hyperparameters increases}

\subsection{Sensitivity Analysis (RQ3)}

\textbf{The tradeoff between hyperparameters need to be significantly considered.} As demonstrated in Figure \ref{fig:4b_ablation}, steering both shallow and deep tokens with $\alpha$ ranging from $3.0$ to $6.0$ and $\beta$ ranging from $0.1$ to $0.9$ can effectively enhance the ASR, though achieving this requires careful calibration to avoid compromising text quality. For example, when $\alpha$ is $4.0$ and $\tau$ is $150$, increasing $\beta$ from $0.3$ to $0.7$ results in the ASR rising from $0.2125$ to $0.8385$. However, maintaining a sufficient steering strength is critical, as a very small $\alpha$ fails to improve the ASR even if $\beta$ is large, with values staying below $0.1$ when $\alpha$ is $3.0$. Ultimately, the selection of these hyperparameters involves a trade-off with linguistic fluency because excessively high values lead to a continuous decrease in the D2G. This reduction in diversity, observed when $\alpha$ is $6.0$ and $\beta$ increases toward $0.9$, reflects an increase in repetitive blocks that makes the resulting jailbreak content unavailable for practical use.

% \textbf{In most of the case, increasing the size of $\mathcal{D}$  can improve ASR.} As shown in Table \ref{table:ablation_prefix}, 

\textbf{In most of the case, increasing the size of $\mathcal{D}$ can improve ASR.} As shown in Table \ref{table:ablation_prefix}, total prefix combinations increasing from 9 to 100 lead to average ASR gains from 0.6716 to 0.8295 for Qwen3, while perform between 0.8295 and 0.8352 for Llama-3.1 with error 0.69\%. These results confirm that larger prefix sets enable more accurate steering of the model's hidden states toward compliant responses by providing a more robust estimation of the boundary between refusal and helpfulness.

\begin{table}[h]
	\centering
	\caption{Comparison for Different Prefix Sets Size}
	\label{table:ablation_prefix}
	\renewcommand{\arraystretch}{1}
    % \begin{tabular}{ccc|ccc|c}
    \begin{tabular}{wc{0.45cm}wc{0.45cm}wc{0.45cm}|wc{1cm}wc{1cm}wc{1cm}|wc{1cm}}
        \toprule[1pt]
        $\mathcal{D}_{com}$ & $\mathcal{D}_{ref}$ & Total & AdvBench & Malicious & Forbidden & Avg \\
        \midrule[1pt]

        \multicolumn{7}{c}{Qwen3-4B-Instruct-2507, $\tau=150$, $\alpha=5.0$, $\beta=0.5$} \\
        
        \midrule[1pt]

        $3$ & $3$ & $9$ & $0.8163$ & $0.7100$ & $0.4885$ & $0.6716$ \\

        $5$ & $5$ & $25$ & $0.8788$ & $0.7750$ & $0.5385$ & $0.7308$ \\

        $8$ & $8$ & $64$ & $0.8923$ & $0.7250$ & $0.5231$ & $0.7135$ \\
        
        $10$ & $10$ & $100$ & $0.9019$ & $0.7950$ & $0.5718$ & $0.7562$ \\

        \midrule[1pt]

        \multicolumn{7}{c}{Llama-3.1-8B-Instruct, $\tau=150$, $\alpha=0.8$, $\beta=0.5$} \\
        \midrule[1pt]

        $3$ & $3$ & $9$ & $0.9904$ & $0.8700$ & $0.6282$ & $0.8295$ \\

        $5$ & $5$ & $25$ & $0.9702$ & $0.9450$ & $0.5949$ & $0.8352$ \\

        $8$ & $8$ & $64$ & $0.9817$ & $0.8700$ & $0.6141$ & $0.8219$ \\
        
        $10$ & $10$ & $100$ & $0.9702$ & $0.9350$ & $0.5833$ & $0.8295$ \\
        
        \bottomrule[1pt]

        % \multicolumn{6}{l}{$^{1}$Abbreviation for MaliciousInstruct and ForbiddenQuestions.}
	\end{tabular}
\end{table}

% ===== ===== ===== ===== ===== ===== ===== ===== ===== ===== ===== ===== =====
\section{Conclusion and Future Works}
In this paper, we introduced ShallowJail to demonstrate that manipulating initial hidden states can effectively bypass LLM safety alignment. Our experiments show that ShallowJail achieves an ASR up to 0.9702 on Llama-3.1-8B while maintaining high linguistic diversity as evidenced by D2G scores that often surpass natural response levels. These findings confirm that safety alignment is critically vulnerable during the generation of the first few shallow tokens. Consequently, this work highlights the urgent need for more robust defense mechanisms that persist throughout the entire generation process. In the future, we will build up the adaptive steering method to trigger the jailbreak.

% ===== ===== ===== ===== ===== ===== ===== ===== ===== ===== ===== ===== =====
\section*{Ethical Consideration}
We conducted all experiments using publicly available datasets and strictly controlled the jailbreak experiments only at local machine. We did not utilize our method to target specific individuals or deployed public or commercial systems.

\section*{Acknowledgment}
The part of sentences in this paper was polished by Google Gemini. The authors wish to thank the reviewers for their helpful comments and suggestions. This work is supported in part by [Anonymous].

% ===== ===== ===== ===== ===== ===== ===== ===== ===== ===== ===== ===== =====
\bibliographystyle{IEEEtran}
\bibliography{./refer.bib}

\end{document}